# Electron-hole asymmetry in the phase diagram of carrier-tuned CsV$_3$Sb$_5$

Andrea N. Capa Salinas,[1] Brenden R. Ortiz,[1] Calvin Bales,[2]

Jonathan Frassineti,[2] Vesna F. Mitrović,[2] and Stephen D. Wilson[1, *]

[1] *Materials Department, University of California, Santa Barbara, California 93106, USA*

[2] *Physics Department, Brown University, Providence, RI 02912, USA*

(Dated: August 1, 2023)

Here we study the effect of electron doping the kagome superconductor CsV$_3$Sb$_5$. Single crystals and powders of CsV$_3$Sb$_{5-x}$Te$_x$ are synthesized and characterized via magnetic susceptibility, nuclear quadrupole resonance, and x-ray diffraction measurements, where we observe a slight suppression of the charge density wave transition temperature and superconducting temperature with the introduction of electron dopants. In contrast to hole-doping, both transitions survive relatively unperturbed up to the solubility limit of Te within the lattice. A comparison is presented between the electronic phase diagrams of electron- and hole-tuned CsV$_3$Sb$_5$.

## I. INTRODUCTION

The interplay between charge density wave (CDW) order and superconductivity (SC) in the $A$V$_3$Sb$_5$ ($A$ = K, Rb, Cs) class of kagome superconductors remains a focus of ongoing research [22, 23, 35]. The band structure of $A$V$_3$Sb$_5$ hosts a series of saddle points near the Fermi level [21], giving rise to Van Hove singularities theorized to promote the formation of unconventional electronic states [8, 9, 11, 32]. At high temperature, nesting effects combined with electron-phonon coupling are proposed to stabilize the formation of a CDW state [13, 28, 33]. At lower temperatures, superconductivity arises from this CDW state, and the coupling between the two phase transitions can provide insights into a number of the proposed instabilities in this class of materials.

Specifically, the coupling between CDW order and SC in $A$V$_3$Sb$_5$ compounds has been experimentally explored through a number of approaches. These include tracking the evolution of both order parameters as the system is perturbed *via* chemical pressure [14–16, 20, 40], change in dimensionality [25, 26, 31], external pressure [1, 2, 5– 7, 29, 30, 36–39, 41], uniaxial strain [24], and chemical doping [4, 16–19, 27, 34]. One function of these perturbations is to shift the chemical potential about the multiple Van Hove singularities nearby, though the dominant perturbation in the case of doping is often considered to be the orbitally-selective modification of the Sb $p_z$ pocket at the $\Gamma$ point in the Brillouin zone [12].

In the case of carrier doping, hole-doping has been shown to rapidly drive the suppression of long-range CDW order as well as an accompanying increase in the SC transition temperature ($T_c$) [18, 19, 34]. For the case of CsV$_3$Sb$_5$, $T_c$ evolves in a nonmonotonic fashion and two SC domes emerge. The second dome appears in the regime where long-range CDW is fully suppressed, and there are qualitative similarities observed in the pressure-tuned phase diagram of CsV$_3$Sb$_5$ [1, 37, 38]. The evolution of charge correlations into an incommensurate, quasi-1D regime beyond the phase boundary of 3D CDW order suggests a link between the formation of two SC domes and a crossover in the character of charge correlations. [7, 10]

One less explored question is whether there exists an electron-hole asymmetry in the carrier-tuned phase diagram of CsV$_3$Sb$_5$. In a rigid band shift model, the relative shift of the Van Hove points relative to $E_F$ should be important to the response of the system, and in the more realistic case of orbitally-selective doping, the impact of the relative changes in the Sb $p_z$ mixed bands on the CDW state stand to inform more about their role in the formation of charge order. Naively, a clean means of introducing holes has been demonstrated via the substitution of Sn onto the Sb sites of $CsV_3Sb_{5-x}Sn_x$ which preserves the core V-based kagome matrix [19]. The electron-doping counterpart to this can be achieved via Te substitution onto the Sb sites in $CsV_3Sb_{5-x}Te_x$, which is the focus of this paper.

*email: stephendwilson@ucsb.edu

Here we present x-ray diffraction, nuclear quadrupole resonance, and susceptibility measurements characterizing the electron-doped phase diagram of $CsV_3Sb_5$. Our data demonstrate a limited solubility of Te into the $CsV_3Sb_{5-x}Te_x$ matrix before phase separation occurs near $x \approx 0.08$ and that Te preferentially occupies the Sb-sites in the V-kagome plane. In contrast to hole-doping, the introduction of electrons results in a relatively weak perturbation to the system— one where the CDW onset temperature is only slightly suppressed and SC is weakly suppressed in a monotonic fashion. The likely dominant driver of the weak suppression of both states is the introduction of disorder via Sb-substitution, establishing a sharp contrast to the hole-doped phase diagram of this system.

## II.     MATERIALS AND METHODS

### A.     Powder and Single Crystal Synthesis

Powders of $CsV_3Sb_{5-x}Te_x$ were synthesized inside a glovebox filled with argon (water and oxygen levels at < 0.5 ppm) by measuring stoichiometric amounts of elemental Cs (solid, Alfa 99.98%), V (powder, Sigma 99.9%, previously purified using a 1:10 ratio of EtOH and HCl), Sb (shot, Alfa 99.999%), and Te (lump, Alfa 99.999%). For each composition of Te doping, 6 g batches of the starting materials were ball-milled inside a tungsten carbide vial for 60 min in a SPEX 8000D high-energy ball mill. The resulting powders were extracted inside a glovebox, and then ground and sieved through a 106 micron sieve. Powders were then placed inside an alumina crucible, sealed inside an argon-filled quartz tube, and annealed at 550 C for 48 hours. A post-anneal grind and sieve was performed followed by a second anneal at 450 C for 12 hours. The final powders are gray and reasonably air stable.

Single crystals of $CsV_3Sb_{5-x}Te_x$ were grown by a self-flux method. Different Te concentrations were targeted by the formula $Cs_{20}V_{15}Sb_{120-x}Te_x$, with $x$ = 7.2,9.6. Elements were weighed inside a glovebox to make 4 g batches of fluxes accordingly, each loaded into tungsten carbide vials, and ball milled for 60 min. Precursors were then extracted, loaded into alumina crucibles and sealed inside carbon-coated quartz tubes. Previous attempts at synthesizing $CsV_3Sb_{5-x}Te_x$ crystals were performed inside sealed steel tubes, but were unsuccessful given that elemental Te corrodes steel. The sealed tubes were heated at 900 C for 12 hours and then cooled to 500 C at 2C/hour. Single crystals were extracted manually using IPA.

### B.     Experimental Details

Powder X-ray diffraction data were collected on a Panalytical Emperyan powder diffractometer. Pawley and Rietveld refinements were performed using the software *TOPAS-Academic* [3]. A tabletop scanning tunneling microscope (SEM) Hitachi TM4000Plus was used to analyze concentrations of Te in single crystal samples. Magnetization data for both powders and crystals were measured inside a Quantum Design Magnetic Properties Measurement System (MPMS) using the vibrating sample mode (VSM) to detect the superconducting transitions under a field of 5 Oe, and to measure the charge density wave transition under 10000 Oe. Low-temperature susceptibility data were collected using a Quantum Design Physical Properties Measurement System (PPMS) with a dilution refrigerator insert and the AC susceptiblity option. Room temperature [121]Sb zero-field nuclear quadrupolar resonance (NQR) measurements were performed using a laboratory-made NMR spectrometer and probe. Quadrupolar lines from the $I = 5/2$ Sb nuclei were collected from Fourier transforms of the spin-echo using the same sequence and approach as that detailed in Ref. [19]. Two distinct Sb chemical sites are present in the unit cell, which we label Sb1 and Sb2, and they generate unique frequencies.

### III.     RESULTS AND DISCUSSION

Powders and single crystals of $CsV_3Sb_{5-x}Te_x$ were synthesized in the composition range $0 \leq x \leq 0.1$. The 300 K structure remains $P6/mmmm$ across this composition range, and the tellurium dopants occupy the Sb1 site in the kagome plane as shown in Figure 1. Demonstrating this, NQR data plotted in Figures 1 (c) and (d) show the preferential chemical shift of only the Sb2 sites at both NQR transitions probed. Using the same reasoning as that presented earlier [19], this indicates that Te preferentially occupies the in-plane Sb1 positions, as only changes to Sb1-Sb2 field gradients are observed and (similar distance) Sb2-Sb2 field gradients are unaffected.

Lattice parameters derived from Pawley refinements of powder x-ray data are shown in Figure 2. The resulting $c/a$ ratio plotted in Figure 2 (a) reveals a continuous decrease up to a concentration of $x \approx 0.07$, and for concentrations greater than $x = 0.08$, impurity peaks are observed in the x-ray powder patterns, shown as '*' in Figure 2 (c). This suggests that the solubility limit is $x \approx 0.07 - 0.08$ of Te within the lattice, and a similar deviation from a linear Vergard-like behavior is suggested in the $a$- and $c$-axis lattice parameters plotted in Figure 2 (b) left side, and right side (though the changes are small and at the boundary of resolution).

Now turning to the characterization of $CsV_3Sb_{5-x}Te_x$ below the solubility limit, magnetization and susceptibility data are plotted in Figure 3. Figure 3 (a) shows low-field susceptibility data characterizing the superconducting transition of polycrystalline $CsV_3Sb_{5-x}Te_x$ up to $x = 0.08$. Susceptibility data collected using a dilution insert (4 K - 80 mK) was normalized to overlapping low-field magnetization-derived susceptibility data above 2 K. Within the uncertainty of this normalization procedure, all specimens show a bulk superconducting transition and the final volume fraction was normalized to 100% for comparison between samples. In contrast to the effect of hole-doping, $T_c$ shows a monotonic and gradual decrease as a function of Te concentration.

Magnetization data collected at higher temperatures plotted in Figure 3 (c) reveal that a well-defined CDW transition remains observable for all compositions up to $x = 0.08$. This strongly contrasts the response observed upon hole-doping, where the introduction of Sn immediately broadens and shifts the CDW transition. Quantifying the shift in the CDW onset temperature, Figure 3 (d) plots the derivative $\delta MT/\delta T$ revealing a smooth shift downward in the CDW temperature upon Te doping with minimal broadening of the CDW anomaly in the magnetization data. This is again distinct from the response driven via hole-substitution where a rapid broadening immediately onsets and the CDW transition vanishes near $x = 0.05$ holes per formula unit.

Our results are summarized in Figure 4 where the electronic phase diagram of hole- versus electron-doped $CsV_3Sb_5$ is plotted (ie. Sn versus Te doped). Both electron and hole dopants result in the suppression of the CDW transition temperature; however the suppression is more rapid for hole-doping and the transition vanishes near $x = 0.05$. In contrast, the suppression of the CDW temperature is more gradual with electron doping, and, crucially, the CDW transition remains well-defined until the solubility limit of Te is reached. The SC transition evolves smoothly downward with electron-doping and does not follow a trivial enhancement via a trade-off in the density of states as the parent CDW state is weakened. This simultaneous suppression of the CDW onset temperature and $T_c$ suggests that disorder introduced via chemical alloying may play a role in the suppression of each phase.

To test this, select single crystals of $CsV_3Sb_{5-x}Te_x$ were measured and their transition temperatures overplotted with those from powder samples in Figure 4. The apparent onset temperature of the CDW state is always higher in powders than crystals, but both crystals and powders show a qualitatively similar smooth decrease in the CDW transition up to the solubility limit. The superconducting $T_c$ in powders, however, is often degraded relative to single crystals due to disorder effects (such as strain and plastic deformation) incurred during powder preparation. As a result, trends in $T_c$ as a function of doping are often more reliable in crystals. Figure 4 shows that the $T_c$ for single crystals of $CsV_3Sb_{5-x}Te_x$ are indeed higher than those of powders and that the suppression of $T_c$ with Te-doping is severely reduced. This supports the notion of a disorder-induced suppression of $T_c$ as a function of impurity concentration rather than electron doping. Notably, the canonical trade-off between suppressed CDW order and enhanced SC due to density of states effects is absent, further supporting the idea of a dominant role of dopant-induced disorder.

The above results are derived using chemical dopants that avoid the V-sites in the kagome network; however qualitatively similar trends in the evolution of CDW and SC order parameters have been reported in individual studies leveraging V-site substitution. Ti-doped $CsV_3Sb_5$ renders a phase diagram qualitatively similar to Sn-doped $CsV_3Sb_5$ [34] and Cr-doped $CsV_3Sb_5$ also reveals an asymmetric persistence of CDW order [4]. We note one difference is that $T_c$ is reported to be rapidly suppressed in Cr-doped samples, which is distinct from our Te-doped data. This is likely due to a stronger impurity potential native to the Cr dopants residing directly within the kagome network.

## IV. CONCLUSIONS

In conclusion, our results illustrate a strong electron-hole asymmetry to the electronic phase diagram of carrier-doped $CsV_3Sb_5$. Electron-doping via Te-substitution in $CsV_3Sb_{5-x}Te_x$ largely preserves the CDW state whereas hole-doping rapidly suppresses long-range CDW order and renormalizes the nature of charge correlations. At lower temperatures, light electron-doping also largely preserves the superconducting state whereas hole-doping creates a nonmonotonic, "double dome" response. We suggest that the slight suppression in the onset temperatures of both the CDW and SC orders observed upon Te-substitution arises from alloy-induced disorder rather than a doping-driven effect and that both transitions are robust to light electron-doping. Our findings motivate a deeper theoretical exploration of electron-hole asymmetries in the carrier-tuned band structure of $CsV_3Sb_5$ and related compounds as a means of isolating the dominant Van Hove points and other band features responsible for driving electronic order in this family of compounds.

## CONFLICT OF INTEREST STATEMENT

The authors declare that the research was conducted in the absence of any commercial or financial relationships that could be construed as a potential conflict of interest.

## AUTHOR CONTRIBUTIONS

A.C.S. and B.R.O. synthesized and characterized the materials. C.B, J.F., and V.F.M. performed NQR measurements. A.C.S. and S.D.W. wrote the manuscript with input from all authors.

**FUNDING**

This work was supported by the US Department of Energy (DOE), Office of Basic Energy Sciences, Division of Materials Sciences and Engineering under Grant No. DE-SC0017752 (B.R.O. and A.C.S.). S.D.W. acknowledges partial support via the UC Santa Barbara NSF Quantum Foundry funded via the Q-AMASE-i program under award DMR-1906325. A.C.S and B.R.O acknowledge use of the shared experimental facilities of the NSF Materials Research Science and Engineering Center at UC Santa Barbara (DMR- 1720256). The UC Santa Barbara MRSEC is a member of the Materials Research Facilities Network (www.mrfn.org). This work was supported in part by U.S. National Science Foundation (NSF) grant No. DMR-1905532 (V.F.M.)

**DATA AVAILABILITY STATEMENT**

The datasets generated in this study can be obtained in DOI: https://doi.org/10.5281/zenodo.8146233.

---

**FIGURE CAPTIONS**

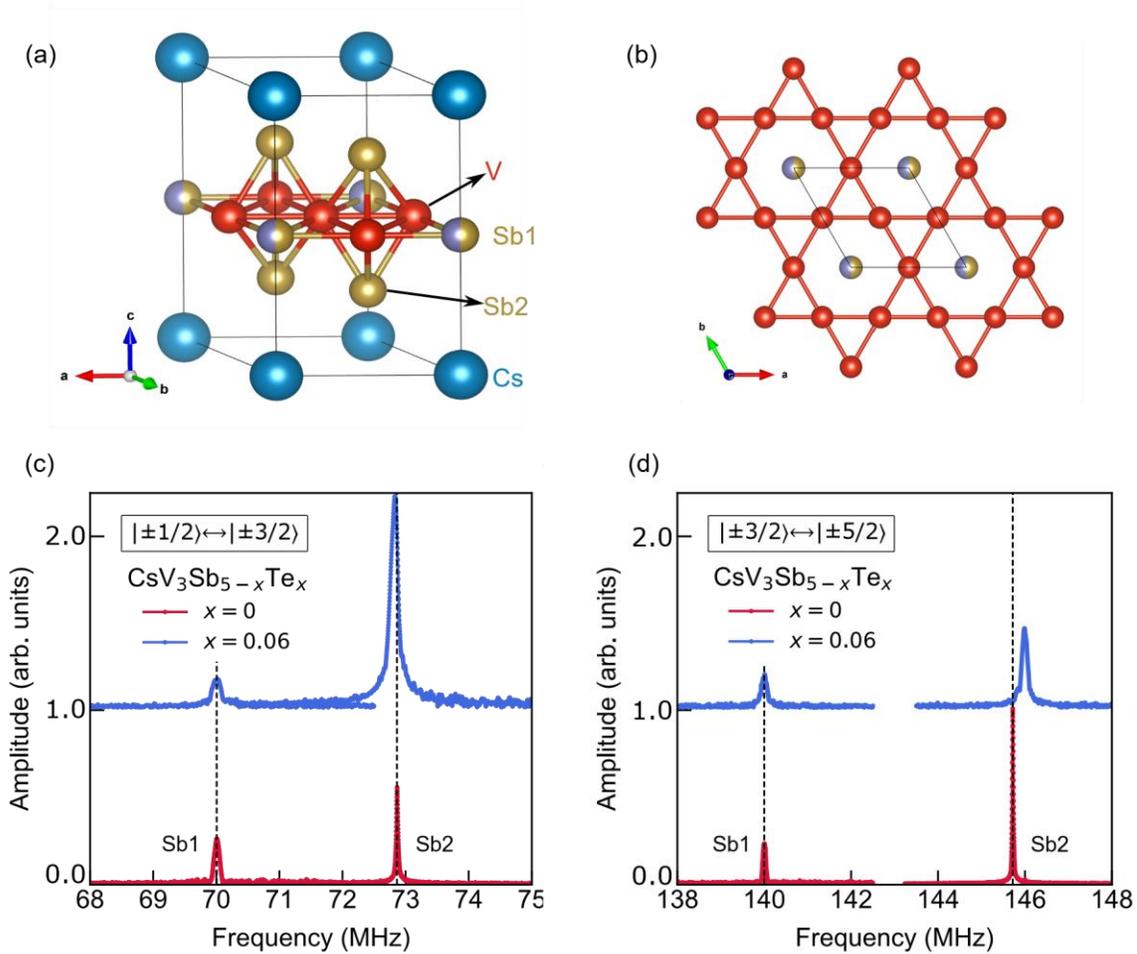

FIG. 1: **(a)** Side and **(b)** top view of $CsV_3Sb_{5-x}Te_x$ structure. NQR data is shown for the **(c)** first transition and **(d)** second transitions of the Sb1 site and Sb2 sites, demonstrating that Te occupies the Sb1 site at $x = 0.06$

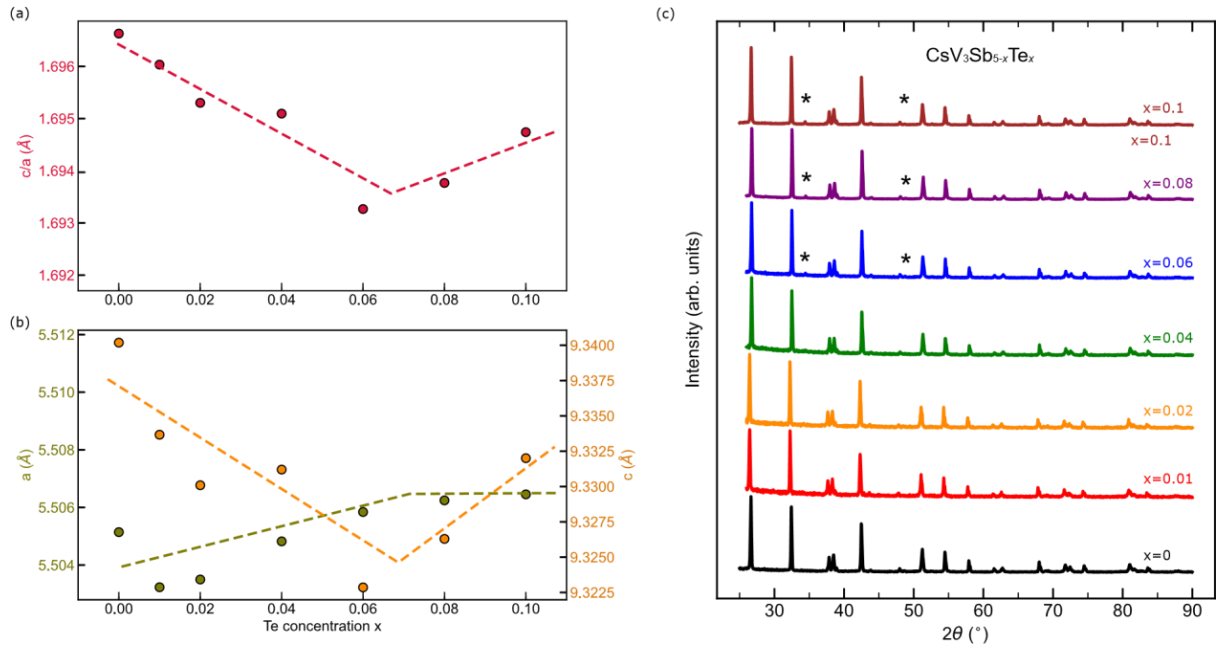

FIG. 2: The CsV$_3$Sb$_{5-x}$Te$_x$ structure does not allow for significant chemical substitution beyond the $x = 0.1$ limit, at which a V-Sb impurity shows up. **(a)** shows the ratio of lattice parameters $c/a$ as a function of $x$ below this limit, and **(b)** shows cell parameters $a$ (left) and $c$ (right) individually as a function of tellurium doping. **(c)** shows the x-ray powder data collected for each concentration $x$ with the onset of an impurity phase marked by * within the patterns

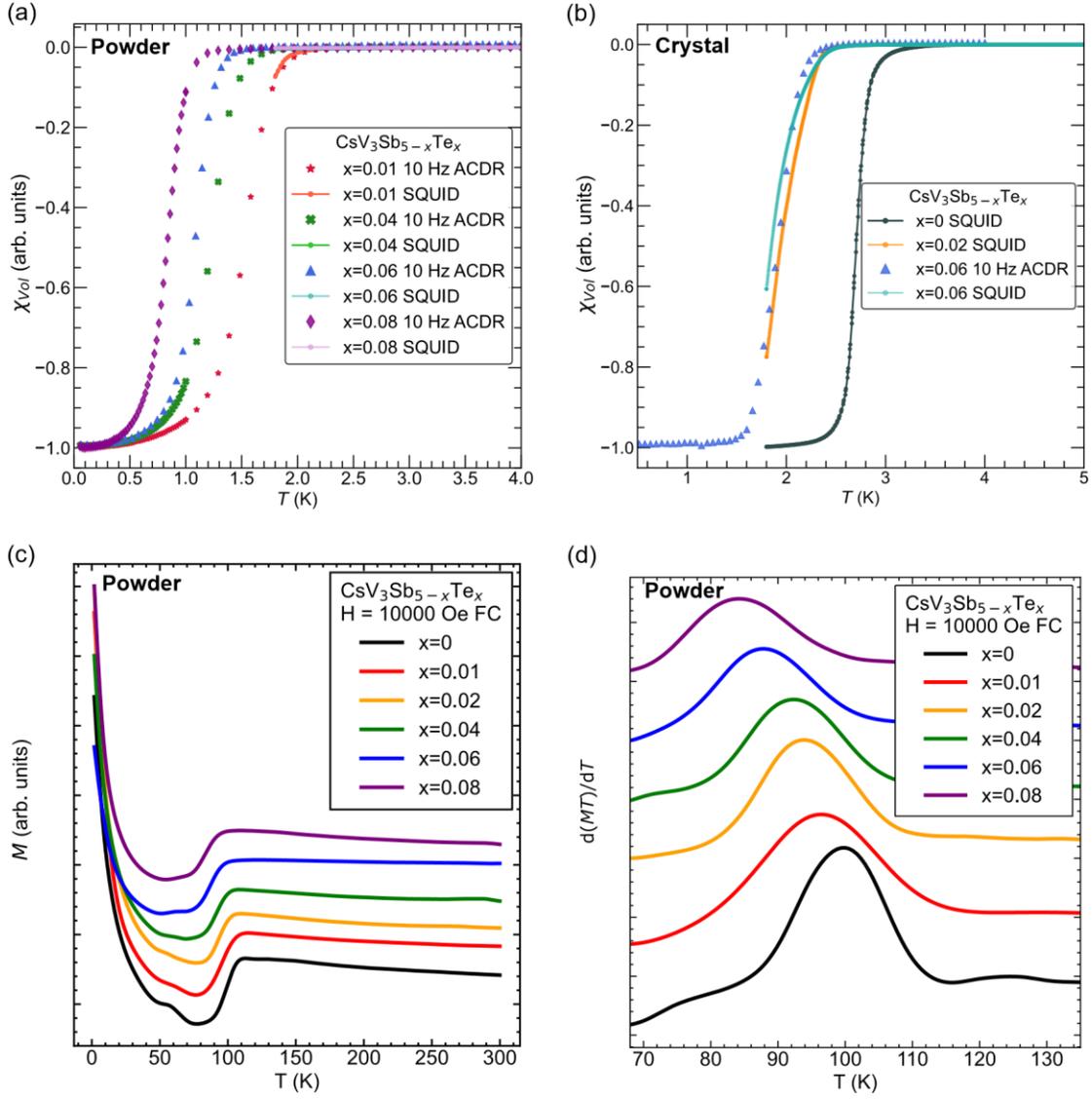

FIG. 3: Superconducting transitions for $CsV_3Sb_{5-x}Te_x$ decrease with increasing Te doping from the parent $T_c$ at 2.5 K, as observed both in powder **(a)** and **(b)** crystal samples. Similarly, the charge density wave transition is weakly suppressed with electron doping, as revealed in **(c)** showing the magnetization plotted as a function of temperature, and more clearly in **(c)** in the plot of $d(MT)/d(T)$.

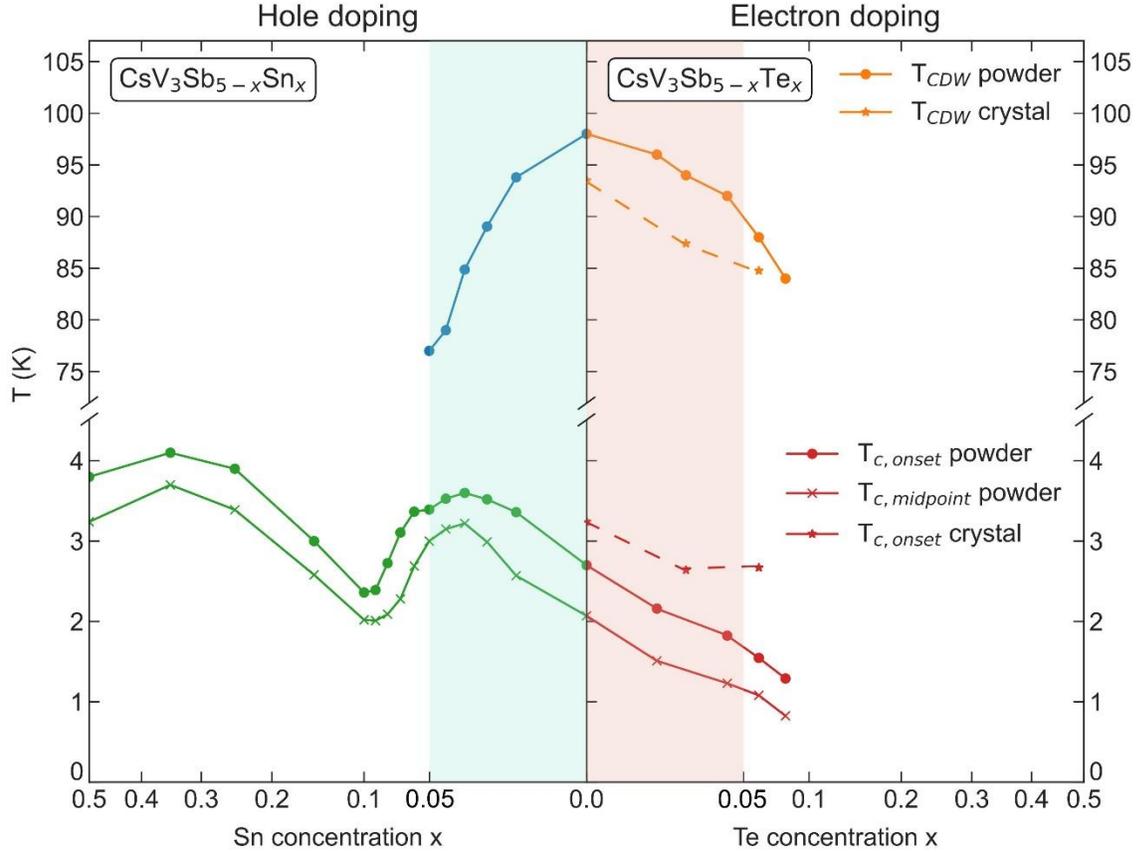

FIG. 4: Plot of the electron-hole asymmetry in the electronic phase diagram of $CsV_3Sb_5$. Electron doping on the Sb site of $CsV_3Sb_5$ shows only a weak suppression of the CDW state, and the CDW state in $CsV_3Sb_{5-x}Te_x$ system persists beyond Te=0.05. The superconducting transition is only weakly suppressed under light electron-doping in $CsV_3Sb_{5-x}Te_x$. This contrasts the dramatic suppression of CDW order and the double-dome evolution of superconductivity that emerges upon hole-doping. Data for hole-doping via Sn substitution was adapted from [19].